\documentclass[sigconf]{acmart}

\usepackage{booktabs} 
\usepackage{subfigure}
\usepackage{amsmath}
\usepackage{multirow}
\usepackage{multicol}
\usepackage{balance}
\usepackage{url}
\makeatletter
\g@addto@macro{\UrlBreaks}{\UrlOrds}
\makeatother
\setcopyright{rightsretained}
\usepackage{amssymb,amsmath,amsthm,enumitem}
\usepackage[font=small,belowskip=-12pt,aboveskip=0pt]{caption}

\acmArticle{4}
\acmPrice{15.00}

\copyrightyear{2019}
\acmYear{2019} 
\setcopyright{iw3c2w3}

\acmConference[WWW '19]{Proceedings of the 2019 World Wide Web Conference}{May 13--17, 2019}{San Francisco, CA, USA}
\acmBooktitle{Proceedings of the 2019 World Wide Web Conference (WWW '19), May 13--17, 2019, San Francisco, CA, USA}
\acmPrice{}

\acmDOI{10.1145/3308558.3313580}

\acmISBN{978-1-4503-6674-8/19/05}
\begin{document}

\setlength{\abovedisplayskip}{2pt}
\setlength{\belowdisplayskip}{2pt}

\title{Recommender Systems with Heterogeneous Side Information}

\author{Tianqiao Liu}
\affiliation{%
  \institution{TAL AI Lab}
  \streetaddress{No.6, Danling Street}
  \city{Beijing, China}
  \postcode{100080}
}
\email{liutianqiao@100tal.com}

\author{Zhiwei Wang}
\authornote{Work was done when the author did internship in TAL AI Lab.}
\affiliation{%
  \institution{Data Science and Engineering Lab}
  \streetaddress{}
  \city{Michigan State University}
  \postcode{}
}
\email{wangzh65@msu.edu}

\author{Jiliang Tang}
\affiliation{%
  \institution{Data Science and Engineering Lab}
  \streetaddress{}
  \city{Michigan State University}
  \postcode{}
}
\email{tangjili@msu.edu}

\author{Songfan Yang}
\affiliation{%
  \institution{TAL AI Lab}
  \streetaddress{No.6, Danling Street}
  \city{Beijing, China}
  \postcode{100080}
}
\email{yangsongfan@100tal.com}

\author{Gale Yan Huang}
\affiliation{%
  \institution{TAL AI Lab}
  \streetaddress{No.6, Danling Street}
  \city{Beijing, China}
  \postcode{100080}
}
\email{galehuang@100tal.com}

\author{Zitao Liu}
\authornote{Corresponding author: Zitao Liu.}
\affiliation{%
  \institution{TAL AI Lab}
  \streetaddress{No.6, Danling Street}
  \city{Beijing, China}
  \postcode{100080}
}
\email{liuzitao@100tal.com}

\begin{abstract}
In modern recommender systems, both users and items are associated with rich side information, which can help understand users and items. Such information is typically heterogeneous and can be roughly categorized into flat and hierarchical side information. While side information has been proved to be valuable, the majority of existing systems have exploited either only flat side information or only hierarchical side information due to the challenges brought by the heterogeneity. In this paper, we investigate the problem of exploiting heterogeneous side information for recommendations. Specifically, we propose a novel framework jointly captures flat and hierarchical side information with mathematical coherence. We demonstrate the effectiveness of the proposed framework via extensive experiments on various real-world datasets. Empirical results show that our approach is able to lead a significant performance gain over the state-of-the-art methods. 
\end{abstract}



\maketitle

\section{Introduction}

Recommender systems can mitigate the information overload problem by providing online users with the most relevant information~\cite{sarwar2001item,su2009survey,ricci2015recommender}. A successful recommender system often requires accurate understanding of user preferences. Collaborative filtering, which models the interactions between users and items, is one of the most popular techniques to achieve this goal~\cite{sarwar2001item,koren2009matrix,he2017neural}. Traditional collaborative filtering based recommender systems have been proven to be suffered from the data sparsity and cold-start problems~\cite{adomavicius2005toward,su2009survey}. On the other hand, in addition to interactions, users and items are often associated with side information, which has become increasingly available in real-world recommender systems~\cite{fang2011matrix,ning2012sparse}. Such side information provides independent sources for recommendations, which can mitigate the data sparsity and cold start problems and have great potentials to boost the performance. As a consequence, a large body of research has been developed to exploit side information for recommendations~\cite{yang2016learning,lu2012exploiting,ricci2015recommender,wang2018exploring}.
\begin{figure}
	\centering
	\includegraphics[scale=0.6]{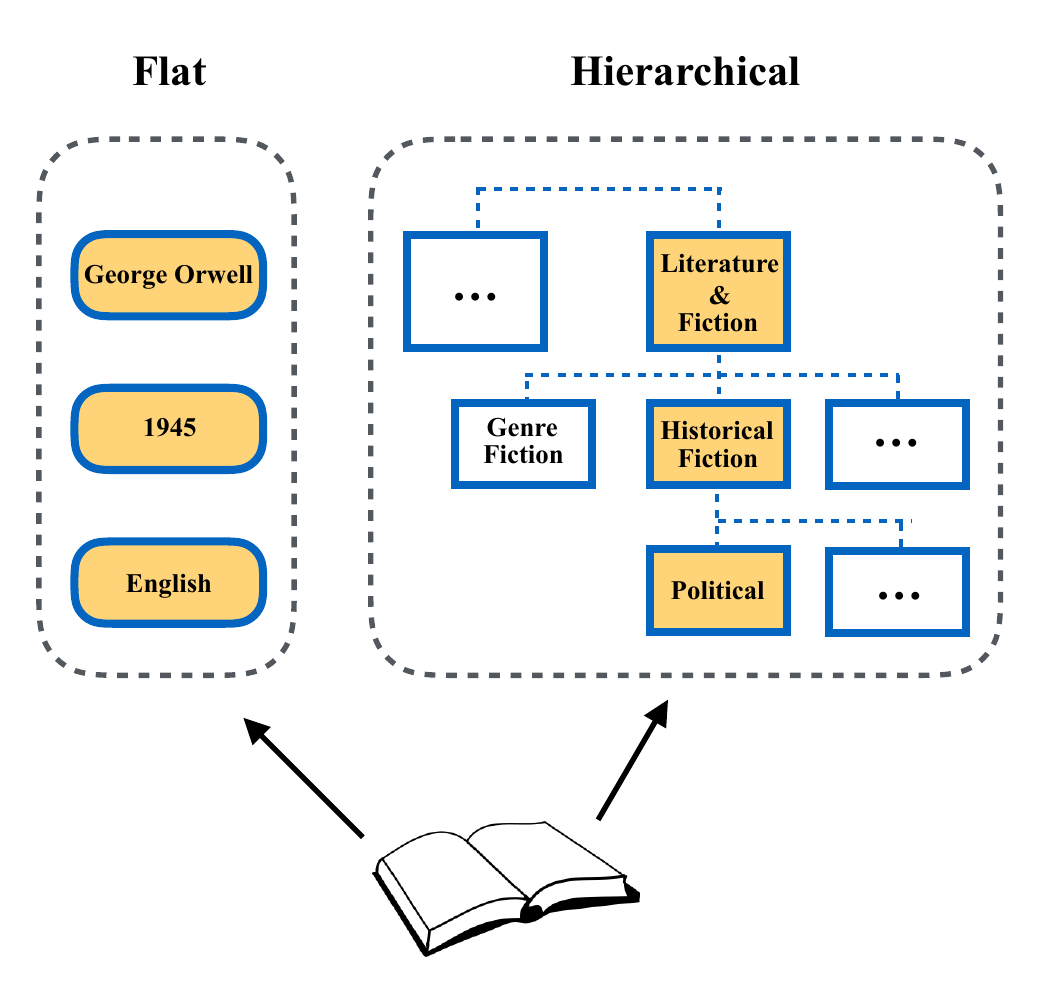}
	\caption{An illustration of flat and hierarchical side information.}
	\label{fig:intro}
\end{figure}
Side information is typically heterogeneous, which can be roughly categorized into flat and hierarchical side information~\cite{wang2018exploring}. Flat and hierarchical side information are referred to attributes associated with users and items presenting no hierarchical and hierarchical structures, respectively~\cite{wang2018exploring}. Take books for example, side information of one book can include the publish year, the book authors, the written language, and the genres it belongs to. Attributes such as year, author and language, presenting no hierarchical structures, are flat. The genres, however, contain {\it subtypeOf} relationship and can be organized in a hierarchical structure. Figure~\ref{fig:intro} gives a concrete example, which shows six attributes of the book {\it Animal Farm} with colored background. The flat information includes {\it George Orwell}, {\it 1945} and {\it English} and are listed in the left part of figure. {\it Literature\&Fiction}, {\it Historical Fiction}, and {\it Political} are the genres {\it Animal Farm} belongs to according to Amazon Web Store. As shown in the figure, these genres are organized into a hierarchical structure as genre$\rightarrow$subgenre$\rightarrow$detailed-genre such that {\it Animal Farm} firstly belongs to the genre {\it Literature\&Fiction}, under which there are sub-genres such as {\it Historical Fiction} and {\it Genre Fiction}. In this sub-genre level, {\it Animal Farm} belongs to {\it Historical Fiction}. It further falls into a more detailed-category {\it Political}. Likewise, users are also associated with both flat information such as age, gender, education level and hierarchical information such as communities they belong to and the places of their birth.

There are numerous works incorporating flat or hierarchical side information for recommendations ~\cite{fang2011matrix,ning2012sparse,yang2016learning,lu2012exploiting,ricci2015recommender,wang2018exploring}.  Most of these systems have been designed to exploit either only flat side information or only hierarchical side information. The major reason is that flat and structured side information are intrinsically different and it is challenging to jointly exploit them. One trivial solution is to flatten the hierarchy and treat it as flat information. However, such solution ignores the unique properties of hierarchical information, which have been proven beneficial by previous works~\cite{lu2012exploiting,wang2018exploring,menon2011response}. In fact, both flat and hierarchical side information can provide valuable information for understanding user preferences and item characteristics. For instance, female are generally more interested in high heel shoes than male and items belonging to the same detailed-genre are likely to be more similar than those in the same sub-genre. Thus, it is desired to design frameworks incorporating the two types of side information simultaneously. 

In this paper, we investigate the problem of exploiting both flat and hierarchical side information for recommendations. We propose a novel framework, which aims to address two challenges:~(1) how to jointly capture heterogeneous side information and~(2) how to mathematically use them for recommendations. The main contributions of our work can be summarized as follows:

\begin{itemize}
	\item We provide a principled approach to simultaneously capture both flat and hierarchical information mathematically.
	\item We introduce a unified recommendation framework HIRE, which can model {\bf H}eterogeneous side {\bf I}nformation for {\bf RE}commendation coherently.
	\item We conduct extensive experiments with various real-world datasets to validate the effectiveness of the proposed framework and understand the importance of flat and hierarchical side information for recommendations.
\end{itemize}


\section{Methodology}
In this section, we present the proposed recommendation framework that coherently captures flat and hierarchical information of both users and items. Specifically, we firstly introduce the notations that will be used in the rest of the paper and then describe a basic model which forms the basis of the framework. After that, we go into details of the framework components that model the flat and heterogeneous information, respectively, combining of which  leads to an optimization problem. Finally, we propose an efficient algorithm to solve it. 

Throughout this paper, regular letters are used to denote scalers. The vectors and matrices are represented by bold lower-case letters such as ${\bf h} $ and bold upper-case letters such as {\bf W}, respectively. In addition, for an arbitrary matrix ${\bf W}$, we use ${\bf W}(i,:)$ and ${\bf W}(:,j)$ to denote the $i^{th}$ row and $j^{th}$ column of it, respectively; and the $(i,j)^{th}$ entry of ${\bf W}$ is represented as ${\bf W}(i,j)$. The transpose and Frobenius norm of ${\bf W}$ is denoted as ${\bf W}^T$ and $\|{\bf W}\|_F^2$, respectively. Moreover, let $\mathcal{U}=\{u_1, u_2, \cdots, u_n\}$ to be the set of $n$ users and $\mathcal{V}=\{v_1, v_2, \cdots, v_m\}$ be the set of $m$ items. We assume there exists a user-item rating matrix ${\bf R}\in \mathbb{R}^{n\times m}$ and if a user $u_i$ has rated item $v_j$, ${\bf R}(i,j)>0$ denotes the rating score, otherwise, ${\bf R}(i,j)=0$. In addition, let ${\bf X}\in \mathbb{R}^{d_x\times n}$ and ${\bf Y}\in \mathbb{R}^{d_y\times m}$ be the matrices that contain flat side information of users and items with $d_x$ and $d_y$ associated attributes, respectively. Next, we will describe a basic recommendation model, based on which we will build the whole framework. 

\vspace{-0.2cm}

\subsection{The Basic Model}
Weighted matrix factorization is an effective approach used in collaborative filtering based recommender systems to obtain users' and items' representations that contain information regarding users' preference and items' characteristics. Specifically, it decomposes the rating matrix and models the users and items in the same low-dimensional latent space. Mathematically, weighted matrix factorization solves the following optimization problem:

\begin{align}
\min_{{\bf U}, {\bf V}} \underbrace{\|{\bf M} \odot ({\bf R} - {\bf U}{\bf V})\|^2_F + \lambda (\|{\bf U}\|^2_F+\|{\bf V}\|^2_F)}_{f({\bf U}, {\bf V})} \nonumber
\end{align}

\noindent where $\odot$ is the Hadamard operation denoting element-wise multiplication. ${\bf M}\in \mathbb{R}^{n\times m}$ is the indication matrix such that ${\bf M}(i,j)=1$ if ${\bf R}(i,j)>0$, otherwise, ${\bf M}(i,j)=0$. The obtained matrices ${\bf U} \in \mathbb{R}^{n\times d}$ and ${\bf V} \in \mathbb{R}^{d\times m}$ are the corresponding representations of users and items in the latent space. Thus, the rating score given by $u_i$ to $v_j$ is approximated by the dot item of their latent representations, that are ${\bf U}(i,:)$ and ${\bf V}(:,j)$, respectively. $\lambda$ is used to control the weight of regularization terms that are adopted to avoid overfitting. One of the most important strength of matrix factorization based model is that it allows to incorporate other information in addition to the ratings~\cite{koren2009matrix}. Next, we will base on the basic weighted matrix factorization model to build our framework.

\vspace{-0.2cm}

\subsection{Capturing The Flat Information}

The side information of items or users are intrinsically heterogeneous, which can be flat and heterogeneous. For example, in Figure~\ref{fig:intro}, a book can have both flat attributes such as author, year, and hierarchical attributes such as genres it belongs to. The difference between different types of side information requires special treatment for each individual. In this subsection, we describe the model component that aims to capture the flat side information. To simplify, we first focus on capturing side information of users and then generalize it to that of items.

In weighted matrix factorization, users' latent representation matrix ${\bf U}$ contains their preference indicated by the rating scores they give to items. The side information, however, provides another independent source from which users' preference could be inferred. For example, it is very likely that a programmer is more interested in a mechanical keyboard than a dancer, which suggests that users' occupation could provide an important indicator whether an item should be recommended or not. In addition, both hidden representation ${\bf U}(i,:)$ and side information ${\bf X}(:,i)$ describe the same user $u_i$. Hence, in the same latent space, ${\bf U}(i,:)$ and ${\bf X}(:,i)$ should be similar. With this intuition, we extend the basic weight matrix factorization model to capture the flat side information contained in ${\bf X}$ as follows:

\vspace{-0.3cm}
\begin{align*}
\min_{{\bf U}, {\bf V}} & f({\bf U},{\bf V}) + \gamma\|{\bf S}^u {\bf U}^T - {\bf X}\|^2_F
\end{align*}

\noindent where ${\bf S}^u \in \mathbb{R}^{d_x\times d}$ is the projection matrix that projects users' hidden representations ${\bf U}$ into the same latent space as ${\bf X}$. The Frobenius norm indicates the distance between the representations of users from two perspectives, which are forced to be close. In this way, the learned users representations ${\bf U}$ also capture flat side information contained in ${\bf X}$.

However, in practice, ${\bf X}$ is usually very sparse. For example, while a user profile could include various types of information, making ${\bf X}$ high-dimensional, many users may only provide a part of the profile, which renders ${\bf X}$ very sparse. To address this issue, we adopt autoencoders, which provide a way to obtain robust feature representations in an unsupervised manner~\cite{bengio2013representation} and have been successfully applied in various tasks such as speech enhancement~\cite{lu2013speech}, natural language generation~\cite{li2015hierarchical} and face alignment~\cite{zhang2014coarse}. In this work, we choose to incorporate marginalized denoising autoencoders (MDA) into the proposed model, as it is much more computationally efficient than others~\cite{chen2012marginalized} and we leave incorporating other autoencoders as one future direction. 

MDA firstly takes the side information ${\bf X}=[{\bf x}_1, {\bf x}_2, \cdots, {\bf x}_n]$ as input and corrupts the features to obtain noising version ${\tilde {\bf x}}_i$ for each user $u_i$. The corruption process can be done in different ways. In this paper, we follow the practice in~\cite{chen2012marginalized} and corrupt features by randomly setting each feature to be $0$ with the probability $p\geq0$. In contrast to traditional stacked denoising autoencoders that have the two-level encoder and decoder structures, in MDA, only one single mapping ${\bf W}^u: \mathbb{R}^{d_x} \rightarrow \mathbb{R}^{d_x}$ is used to reconstruct the original features and the reconstruction loss is defined as follows:

\vspace{-0.3cm}
\begin{equation}
\mathcal{L}({\bf W}^u) = \frac{1}{2n}\sum_{i=1}^n\|{\bf x}_i - {\bf W}^u{\tilde {\bf x}}_i\|^2
\end{equation}

The random corruption of the features may lead to the solution ${\bf W}^u$ of high variance. In order to avoid this, $k$-times corruption is performed and the overall reconstruction loss becomes:

\vspace{-0.3cm}
\begin{equation}
\mathcal{L}({\bf W}^u) = \frac{1}{2nk}\sum_{j=1}^k\sum_{i=1}^n\|{\bf x}_i - {\bf W}^u{\tilde {\bf x}}_{i,j}\|^2
\end{equation}

\noindent where ${\tilde {\bf x}}_{i,j}$ denotes the corrupted version of feature ${\bf x}_i$ at the $j^{th}$-time. Written in matrix form, the above loss function can be expressed as:

\vspace{-0.3cm}
\begin{align}
\label{eq:flat_loss}
\mathcal{L}({\bf W}^u) = \frac{1}{2nk}\|{\bar {\bf X}} - {\bf W}^u{\tilde {\bf X}}\|_F^2
\end{align}

\noindent where ${\bar {\bf X}} = [{\bf X},  {\bf X}, \cdots, {\bf X}] \in \mathbb{R}^{d_x \times nk}$ and ${\tilde {\bf X}} = [{\tilde {\bf X}}_1,  {\tilde {\bf X}}_2, \cdots, {\tilde {\bf X}}_k] \in \mathbb{R}^{d_x \times nk}$ with ${\tilde {\bf X}}_j$ denoting the $j^{th}$ corrupted version of the original features ${\bf X}$. The solution of the minimization of the loss function defined in Eq~(\ref{eq:flat_loss}) can be written in a closed form: $ {\bf W}^u = {\bf K}{\bf J}^{-1}$, where ${\bf K} = {\bar {\bf X}}{\tilde{\bf X}}^T$ and ${\bf J}={\tilde {\bf X}}{\tilde {\bf X}}^T$. Ideally, we would like to make infinitely corrupted versions of ${\bf X}$ to obtain the most stable mapping ${\bf W}^u$. This can be achieved by letting $k \rightarrow \infty$ and computing the expectations of ${\bf K}$ and ${\bf J}$~\cite{chen2012marginalized}. 

With the obtained mapping layer ${\bf W}^u$, robust features of users can be easily constructed from original feature matrix ${\bf X}$ by ${\bf W}^u{\bf X}$. Hence, the framework that captures the flat side information of users with robust feature mapping becomes:

\begin{equation}
\min_{{\bf U}, {\bf V}, {\bf S}^u, {\bf W}^u}  f({\bf U}, {\bf V}) + \gamma(\|{\bf S}^u {\bf U}^T - {\bf W}^u{\bf X}\|^2_F + \|{\bar {\bf X}} - {\bf W}^u{\tilde {\bf X}}\|_F^2)
\end{equation}

Similarly, the flat information of items can also be captured as follows:


\begin{equation}
\min_{{\bf U}, {\bf V}, {\bf S}^v, {\bf W}^v} f({\bf U}, {\bf V}) + \theta(\|{\bf S}^v {\bf V} - {\bf W}^v{\bf Y}\|^2_F + \|{\bar {\bf Y}} - {\bf W}^v{\tilde {\bf Y}}\|_F^2)
\end{equation}

\noindent where ${\bf S}^v \in \mathbb{R}^{d_y \times d}$ is the project matrix that projects ${\bf V}$ into the same space as ${\bf Y}$, ${\bf W}^v \in \mathbb{R}^{d_y \times d_y}$ is the mapping layer that obtains robust features from ${\bf Y}$, ${\bar {\bf Y}} = [{\bf Y},  {\bf Y}, \cdots, {\bf Y}] \in \mathbb{R}^{d_y \times mk}$ and ${\tilde {\bf Y}} = [{\tilde {\bf Y}}_1,  {\tilde {\bf Y}}_2, \cdots, {\tilde {\bf Y}}_k] \in \mathbb{R}^{d_y \times mk}$ with ${\tilde {\bf Y}}_j$ representing the $j^{th}$ corrupted version of ${\bf Y}$.

\vspace{-0.2cm}

\subsection{Incorporating The Hierarchical Information}
Unlike flat information, features in hierarchical information are structured. For example, as shown in Figure~\ref{fig:intro}, the genres of books can be organized into a hierarchical structure. It is very likely that books belong to the detailed genres are more similar than those in sub-genres. Thus, it should be desirable to recommend the book that is in the same detailed-genre with the one that has received high rating score from the user. As hierarchical information is intrinsically different with flat information, the approach introduced in the previous subsection is not suitable. Hence, in this subsection, we introduce how to incorporate hierarchical information by extending the basic matrix factorization model. Without the loss of generality, we firstly introduce the approach to incorporate hierarchical side information of items, which can be naturally applied to that of users.
\begin{figure}
    \centering
    \captionsetup{belowskip=-10pt}
	\includegraphics[scale=0.5]{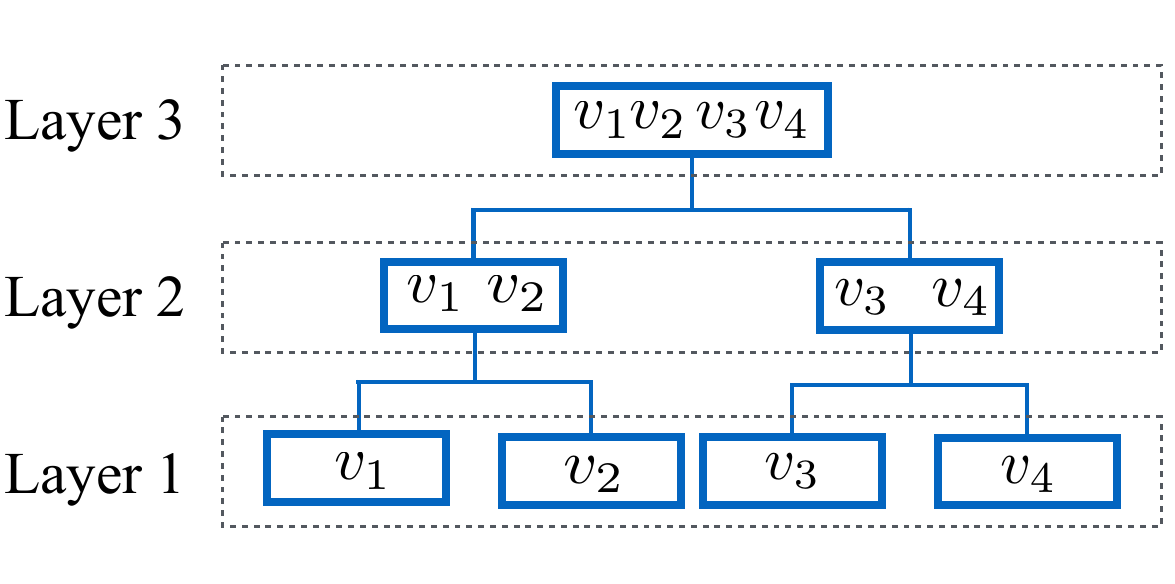}
	\caption{An illustrative example of hierarchical structure.}
	\label{fig:tree}
\end{figure}
Typically, we can use a tree to represent a hierarchical structure. In a tree, each parent node can have different numbers of child nodes, which can also be the parents of the nodes in the next layer. Recall the example given by Figure~\ref{fig:intro}, which shows the hierarchical structure of genres of a book. The genre {\it Literature \& Fiction} is the parent node of several child nodes, such as ${\it Genre Fiction}$, {\it Historical Fiction}, etc. The child node ${\it Historical Fiction}$ is also the parent of other nodes such as ${\it Political}$. The leaf nodes are those who have no child such as ${\it Political}$. From this example, it is easily observed that the hierarchical structure can naturally be characterized by the {\it parent-child} relation. With this intuition, next we introduce how to incorporate the structure information from {\it parent-child} perspective.

The item characteristic matrix ${\bf V} \in \mathbb{R}^{d\times m}$ shows the latent representations of items in a $d$-dimensional latent space. To incorporate the hierarchical structure, we can further decompose ${\bf V}$ into two matrices ${\bf V}^1 \in \mathbb{R}^{m_1\times m}$ and ${\bf V}^{2\prime} \in \mathbb{R}^{d\times m_1}$ such that ${\bf V} \approx {\bf V}^{2\prime} {\bf V}^1$, assuming there are $m_1$ nodes (or sub-categories) in the second layer. Hence, ${\bf V}^1$ indicates the {\it parent-child} relation between the $m_1$ categories in the second layer and $m$ items in the first layer. Moreover, ${\bf V}^{2\prime}$ gives the latent representations of the $m_1$ categories. In this way, the latent representation of $v_j$ can be expressed by the $j^{th}$ item's parent categories and their latent representations as ${\bf V}^{2\prime}{\bf V}^1(:,j)$. If the structured information has more than two layers, we can further decompose ${\bf V}^{2\prime}$ such that ${\bf V}^{2\prime} \approx {\bf V}^{3\prime} {\bf V}^2$, where ${\bf V}^2 \in \mathbb{R}^{m_2\times m_1}$ denotes the {\it parent-child} relation between categories in the third and second layers, respectively. Similarly, ${\bf V}^{3\prime} \in \mathbb{R}^{d\times m_3}$ denotes the representations of $m_2$ categories in the third layer. With this, $j^{th}$ item's representation can be expressed by ${\bf V}^{3\prime}{\bf V}^{2}{\bf V}^1(:,j)$.

The above process can be repeated $q$-$1$ times to capture hierarchical structure of $q$ layers as follows:
\begin{align}
\label{eq:implicit}
{\bf V} \approx {\bf V}^q\cdots{\bf V}^{3} {\bf V}^2{\bf V}^1
\end{align}
where ${\bf V}^i \in \mathbb{R}^{m_i \times m_{i-1}}(1\leq i < q)$, ${\bf V}^q \in \mathbb{R}^{d \times m_{q-1}}$ and $m_0 = m$.

The {\it parent-child} relations indicated by Eq.~(\ref{eq:implicit}) are implicit and they should be in conformity with explicit ones suggested by hierarchical side information. To achieve this, next we extend Eq.~(\ref{eq:implicit}) to incorporate the structures in the side information. Let ${\bf T}^k \in \mathbb{R}^{m_{k-1} \times m_k}$ indicate the {\it parent-child} relation between categories in layer $k$ and layer $k+1$ of the hierarchical structure of side information. Specifically, ${\bf T}^k(i,j)=1$ denotes that the $i^{th}$ category in layer $k$ is the child of $j^{th}$ category in layer $k+1$. Figure~\ref{fig:tree} gives an example, where ${\bf T}^2 \in \mathbb{R}^{2 \times 1}$ and ${\bf T}^2(1,1)=1, {\bf T}^2(2,1)=1$. Thus, the hierarchical structure can be defined by the set $\mathcal{T}=\{{\bf T}^1, {\bf T}^2, \cdots, {\bf T}^q\}$, assuming there are $q$ layers. Intuitively, the presentation of a parent category should be aggregated from those of all the child categories it contains. Thus, a natural way to capture the {\it parent-child} relations in $\mathcal{T}$ is to make parent representations denoted as ${\bf V}^q\cdots{\bf V}^{i}$ to be close to the aggregation of their children's representations denoted as ${\bf V}^q\cdots{\bf V}^{i-1}$. 

In this work, we choose the mean function to be the aggregation function due to its computational efficiency and leave exploring other choices as one future work. Thus, the structure indicated by the item side information can be captured as follows: 
\begin{align*}
\min  \underbrace{\sum_{i=2}^q\|{\bf V}^q\cdots{\bf V}^{i} - {\bf V}^q \cdots {\bf V}^{i-1}{\bf Q}^{i-1}\|_F^2}_{f^v({\bf V}^i,{\bf Q}^{i-1})}
\end{align*}
where ${\bf Q}^k$ is the normalized version of ${\bf T}^k$ and ${\bf Q}^k(i,j) = \frac{{\bf T}^k(i,j)}{\sum_{i=1}^{m_{k-1}}{\bf T}^k(i,j)}$.

In a similar way, we can also incorporate hierarchical information of users as:

\begin{align*}
\min \underbrace{\sum_{i=2}^p\|{\bf U}^i\cdots{\bf U}^{p} - {\bf P}^{i-1}{\bf U}^{i-1} \cdots {\bf U}^{p}\|_F^2}_{f^u ({\bf U}^i, {\bf P}^{i-1})}
\end{align*}
\noindent where ${\bf U}^i \in \mathbb{R}^{n_{i-1} \times n_i}$ and ${\bf P}^k(i,j) = \frac{{\bf C}^k(i,j)}{\sum_{i=1}^{n_{k-1}}{\bf C}^k(i,j)}$ is the normalized version of ${\bf C}^k \in \mathbb{R}^{n_{k-1} \times n_k}$, which indicates the {\it parent-child} relationship in hierarchical side information of users. $p$ is number of layers of the hierarchy and $1\leq i < p$.

\subsection{The Proposed Framework HIRE}
Previous subsections introduce the model components that aim to capture both flat and hierarchical side information. Combining them, the framework HIRE is to solve the following optimization problem:


\begin{align}
\label{eq:optim}
 \min_{\substack{{\bf U}^1,\cdots, {\bf U}^p, {\bf S}^u, {\bf W}^u\\ {\bf V}^1,\cdots, {\bf V}^q, {\bf S}^v, {\bf W}^v}}
&f({\bf U}, {\bf V}) + \alpha f^u ({\bf U}^i, {\bf P}^{i-1}) + \beta f^v({\bf V}^i,{\bf Q}^{i-1}) \\ \nonumber
&+ \gamma(\|{\bf S}^u {\bf U}^T - {\bf W}^u{\bf X}\|^2_F + \|{\bar {\bf X}} - {\bf W}^u{\tilde {\bf X}}\|_F^2) \\ \nonumber
& + \theta(\|{\bf S}^v {\bf V} - {\bf W}^v{\bf Y}\|^2_F + \|{\bar {\bf Y}} - {\bf W}^v{\tilde {\bf Y}}\|_F^2) 
\end{align}

\noindent where ${\bf U} = {\bf U}^1{\bf U}^2\cdots{\bf U}^p$ and ${\bf V} = {\bf V}^q\cdots{\bf V}^2{\bf V}^1$. $\gamma$ and $\theta$ control the contribution of flat information, $\alpha$ and $\beta$ decides the contribution of hierarchical information. Hence, the proposed framework simultaneously models both flat and hierarchical side information with mathematical coherence. Following the tradition~\cite{koren2009matrix}, we will use the gradient descent method to optimize the formulation of the proposed framework. Next we will use ${\bf U}^i$ as one example to illustrate how to get the gradient of parameters due to the page limitation. Before calculating the gradient, we define ${\bf A}^i$, ${\bf H}^i$, ${\bf D}^i$, ${\bf G}_i^k$ and ${\bf B}_i^k$ that can be used to simplify the expressions:
\begin{align*}
&{\bf A}^i = \begin{cases}
{\bf U}^1{\bf U}^2 \ldots {\bf U}^{i-1}, & \mbox{if } i \neq 1 \\
{\bf I}, & \mbox{if } i = 1
\end{cases}\\
&{\bf H}^i = \begin{cases}
{\bf U}^{i+1} \ldots {\bf U}^{p}{\bf V}^q \ldots {\bf V}^1, & \mbox{if } i \neq p \\
{\bf V}^q \ldots {\bf V}^1, & \mbox{if } i = p
\end{cases}\\
&{\bf D}^i = \begin{cases}
{\bf U}^{i+1}{\bf U}^{i+2} \ldots {\bf U}^{p}, & \mbox{if } i \neq p \\
{\bf I}, & \mbox{if } i = p
\end{cases}\\
&{\bf G}_i^k = \begin{cases}
{\bf U}^{k+1} \cdots {\bf U}^{i-1}, & \mbox{if } k \neq i-1 \\
{\bf I}, & \mbox{if } k = i-1
\end{cases}\\
&{\bf B}_i^k =  {\bf P}^k{\bf U}^k{\bf G}_i^k
\end{align*}
where $1 \leq i \leq p$ and $1\leq k < i-1$. By dropping irrelevant terms in Eq.~(\ref{eq:optim}), remaining terms related to ${\bf U}^i$ are as follows:
\begin{align*}
&\mathcal{L}({\bf U}^i) =  \| {\bf M} \odot ({\bf R}-{\bf A}^i{\bf U}^i{\bf H}^i) \|_F^2 + \lambda \| {\bf U}^i\|_F^2 + \alpha  \| ({\bf I}-{\bf P}^i{\bf U}^i){\bf D}^i \|_F^2 \\\nonumber
& +\alpha \sum_{k=1}^{i-1} \| ({\bf G}_i^k - {\bf B}_i^k){\bf U}^i{\bf D}^i \|_F^2  + \gamma \| {\bf S}^u({\bf A}^i{\bf U}^i{\bf D}^i)^T - {\bf W}^u{\bf X}\|_F^2
\end{align*}

Now, we can obtain the gradient of ${\bf U}^i$ as:
\begin{align}
& \frac{\partial \mathcal{L}({\bf U}^i)}{\partial {\bf U}^i} = 
 {{\bf A}^i}^T[{\bf M} \odot ({\bf A}^i{\bf U}^i{\bf H}^i-{\bf R})]{{\bf H}^i}^T + \lambda {\bf U}^i \\ \nonumber
& + \alpha \sum_{k=1}^{i-1}({\bf G}_i^k-{\bf B}_i^k)^T({\bf G}_i^k-{\bf B}_i^k) {\bf U}^i{\bf D}^i{{\bf D}^i}^T\\ \nonumber
&+ \alpha {{\bf P}^i}^T({\bf P}^i{\bf U}^i{\bf D}^i-{\bf D}^i){{\bf D}^i}^T + \gamma {{\bf A}^i}^T({\bf A}^i{\bf U}^i{\bf D}^i{{\bf S}^u}^T - {\bf X}^T{\bf W}^u){\bf S}^u{{\bf D}^i}^T
\end{align}

\subsubsection{Time Complexity Analysis} 

The most expensive operations in the optimization process are updating ${\bf U}^i$ and ${\bf V}^i$, which in each iteration will cost $\mathcal{O}\big(nn_1n_{i-1}  + (2n_in_{i+1} + mm_1)d + n_{i-1}^2(n_i + \sum_{k=1}^{i-1}n_k)\big)$ and $\mathcal{O}\big(mm_1m_{i-1}  + (2m_im_{i+1} + nn_1)d + m_{i-1}^2(m_i + \sum_{k=1}^{i-1}m_k) \big)$, respectively. Thus, assume $N$ iterations are needed in total, the overall time complexity of the optimization process is $\mathcal{O}\Big(\sum_i^p \big ( nn_1n_{i-1}  + (2n_in_{i+1} + mm_1)d + n_{i-1}^2(n_i + \sum_{k=1}^{i-1}n_k) \big) +  \sum_i^q \big (mm_1m_{i-1}  + (2m_im_{i+1} + nn_1)d + m_{i-1}^2(m_i + \sum_{k=1}^{i-1}m_k) \big)\Big)$. 

\section{Experiments}
\begin{table*}
	\begin{center}	
		\caption{Recommendation performance comparison. All prediction differences between HIRE and other methods are statistically significant.}
		\label{table:comparison_result}
		\begin{tabular} { c c c c c c c c c c c c c c}
			\hline 
			\multirow{ 2}{*}{Methods} & \multicolumn{3}{c}{MovieLens~(100K)} & \multicolumn{3}{c}{MovieLens~(1M)} & \multicolumn{3}{c}{BookCrossing} \\
			& 40\% &  60\% & 80\% & 40\% &  60\% & 80\% & 40\% &  60\% & 80\%  \\
			\hline		
			SVD & 1.0152 & 0.9704 &  0.9491 & 0.9161 &  0.9087 & 0.8947 &4.7746&2.8866&2.0899\\ [0.7ex]		
			NMF & 1.0352 & 0.9955 &  0.9715 & 0.9446 &  0.9293 & 0.9227 &2.9381&2.7832&2.6055\\ [0.7ex]
			I-CF & 1.0601 & 1.0485 & 1.0343 &  1.0229 &  1.0065 &  0.9975 & 2.0216 & 1.9989 & 2.2250 \\ [0.7ex]
			NeuMF & 1.0928 &  1.0877 &  1.0849 & 0.9872 &  0.9834 &  0.9825 & 2.0191 &  1.8708 & 1.8586 \\ [0.7ex]
			mSDA-CF & 1.0968 &  1.0891 &  1.0792 &  1.0498 &  1.0482 &  1.0466 & 3.0015 & 2.1992 &1.8692 \\ [0.7ex]
			HSR & 0.9879  &  0.9647  & 0.9376  &  0.9074  &  0.8906 &  0.8742  & 4.8821 &  4.1072 &  3.6137 \\ [0.7ex]
			HIRE & 0.9703 & 0.9398 & 0.9243 &  0.8957 & 0.8778 & 0.8607 & 2.3364 & 1.9193 & 1.8432 \\ [0.7ex]
			\hline   
		\end{tabular}
	\end{center}
\end{table*}

In this section, we firstly introduce the experimental settings. Then, we compare the proposed framework HIRE with representative baselines to answer the first question. Finally, we analyze each model component, which gives answer to the section question. To encourage the reproducible results, we make our code publicly available at: \url{https://github.com/tal-ai/Recommender-Systems-with-Heterogeneous-Side-Information}.

\vspace{-0.2cm}
\subsection{Experimental Settings}
We evaluate the proposed framework on three benchmark datasets MovieLens~(100K), MovieLens~(1M), and BookCrossing and all of them are publicly available~\cite{harper2016movielens,ziegler2005improving}. 
\begin{itemize}
	\item MovieLens~(100K) and MovieLens~(1M) are collected from a movie review website\footnote{https://movielens.org/} where users can give movie rating scores on a scale from 1-5. MovieLens~(100k) contains 100,000 ratings from 1000 users on 1700 movies and MovieLens~(1M) contains 1 million ratings from 6000 users on 4000 movies. For movies, we use genres as hierarchical information; for users, we use age and gender as flat information and occupation as the hierarchical information.
	\item  BookCrossing is a book rating dataset collected from Book-Crossing\footnote{http://www.bookcrossing.com/} community and the rating score is from 1 to 10. After basic data cleaning, we get 17028 ratings from 1009 users on 1816 books. For books, we use publish year and author as flat information and publisher as hierarchical information; for users, we use age and location as the flat and hierarchical information, respectively.
\end{itemize}
For each dataset, we split it into training and test sets such that training set contains $x\%$ of the data and test contains $1-x\%$. We vary $x$ as $\{40, 60, 80\}$. We choose the commonly used Root Mean Square Error~(RMSE) as the measurement metrics of the recommendation performance and lower value of RMSE indicates better performance. In fact, a small improvement in RMSE means a significant improvement of recommender systems~\cite{koren2008factorization}.

\vspace{-0.2cm}
\subsection{Recommendation Performance Comparison}
In this subsection, we evaluate the recommendation performance of proposed framework by comparing it with following representative baselines: 
\begin{itemize}{}
	\item {\bf SVD~\cite{golub1970singular}:} It is a matrix factorization technique that factorizes a user-item rating matrix to obtain latent representations of customers and products via singular-value decomposition~(SVD). In this method, only rating information is used;
	\item {\bf NMF~\cite{gu2010collaborative}:} Non-negative matrix factorization~(NMF) is one of the most popular algorithms used in recommender systems. Unlike SVD, it adds non-negative constraints to the latent representations. This method also only uses rating information;
	\item {\bf I-CF~\cite{sarwar2001item}:} This is a item-based collaborative filtering approach that recommends items to users based on the similarity computed from the rating matrix;
    \item {\bf NeuMF~\cite{he2017neural}:} NeuMF replaces inner product by combining GMF and MLP neural architectures with sharing embedding layer and is able to significantly improves recommendation performance. In this method, only rating information is used.
	\item {\bf mSDA-CF~\cite{li2015deep}:} This method integrates matrix factorization and deep feature learning and achieves state-of-the-art performance. It uses both rating and flat side information and ignores hierarchical one.
	\item {\bf HSR~\cite{wang2018exploring}:} HSR is a state-of-the-art algorithm which is able to capture both rating and structured side information. However, flat information is ignored in this method.
\end{itemize}

Note that the parameters of all methods are selected through five-fold cross validation and the details of parameter selection of the proposed framework are discussed in the later subsections. we repeat each experiment five times and report the average performance in Table~\ref{table:comparison_result}. The following observations can be made from the table:
\begin{itemize}
	\item NeuMF is likely to outperform other traditional CF methods, which suggests the power of deep learning in recommendations. Currently our basic model is based on matrix factorization and it has great potential to choose NeuMF as the basic model to further improve the performance. 
    \item Systems incorporating side information tend to obtain better performance compared to their corresponding systems without side information. This observation supports the importance of side information.
	\item The proposed framework HIRE achieves the best performance in most of the cases. We contribute the superior performance to its ability to capture both flat and structured side information. More details regarding the contribution of each component will be discussed in following subsection.
\end{itemize}
With the above observation, we are able to draw a conclusion to answer the first question: the proposed framework that incorporates heterogeneous side information significantly improves the recommendation performance over the state-of-the-art methods. In the next subsection, we will give a detailed analysis of the contribution from flat and hierarchical side information, respectively.

\subsection{Component Analysis} 
In this subsection, we systematically examine the effect of key components by constructing following model variants:

\begin{itemize}
	\item HIRE-FU: it eliminates the contribution of flat side information of users by setting $\gamma = 0$ in Eq.~(\ref{eq:optim}).
	\item HIRE-FV: it eliminates the contribution of flat side information of items by setting $\theta=0$ in Eq.~(\ref{eq:optim}).
	\item HIRE-SU: it eliminates the contribution of hierarchical side information of users by setting $\alpha=0$ in Eq.~(\ref{eq:optim}).
	\item HIRE-SV: it eliminates the contribution of hierarchical side information of items by setting $\beta=0$ in Eq.~(\ref{eq:optim}).
\end{itemize}

\begin{figure}
	\centering
	\includegraphics[scale=0.5]{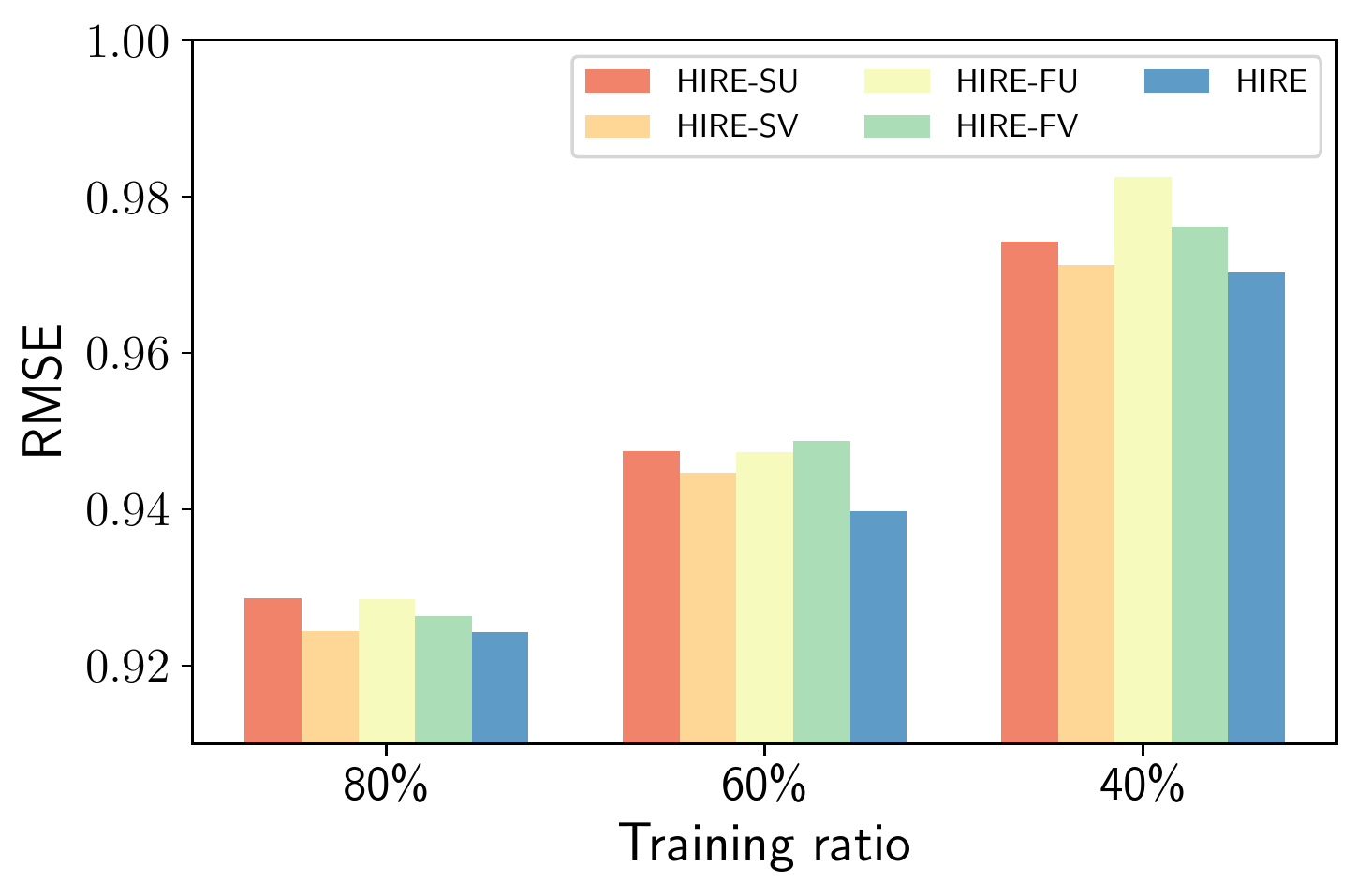}
	\caption{Performance analysis of HIRE with different components.}
	\label{fig:component}
\end{figure}

The recommendation performance on MovieLens~(100K) is shown in Figure~\ref{fig:component}. Since we observe similar results on other datasets, we only show that on MovieLens~(100K) dataset to save space. From the Figure~\ref{fig:component}, we can easily observe that HIRE obtains the least RMSE error among all its variants in all cases. This suggests that recommendation performance degrades when ignoring any type of side information. Thus, it is important to incorporate heterogeneous side information in recommender systems. 

\subsection{Parameter Analysis} 
\begin{figure}
	\begin{center}
		\subfigure[Performance with $\alpha$ abd $\beta$]{\label{fig:intro_1}\includegraphics[scale=0.37]{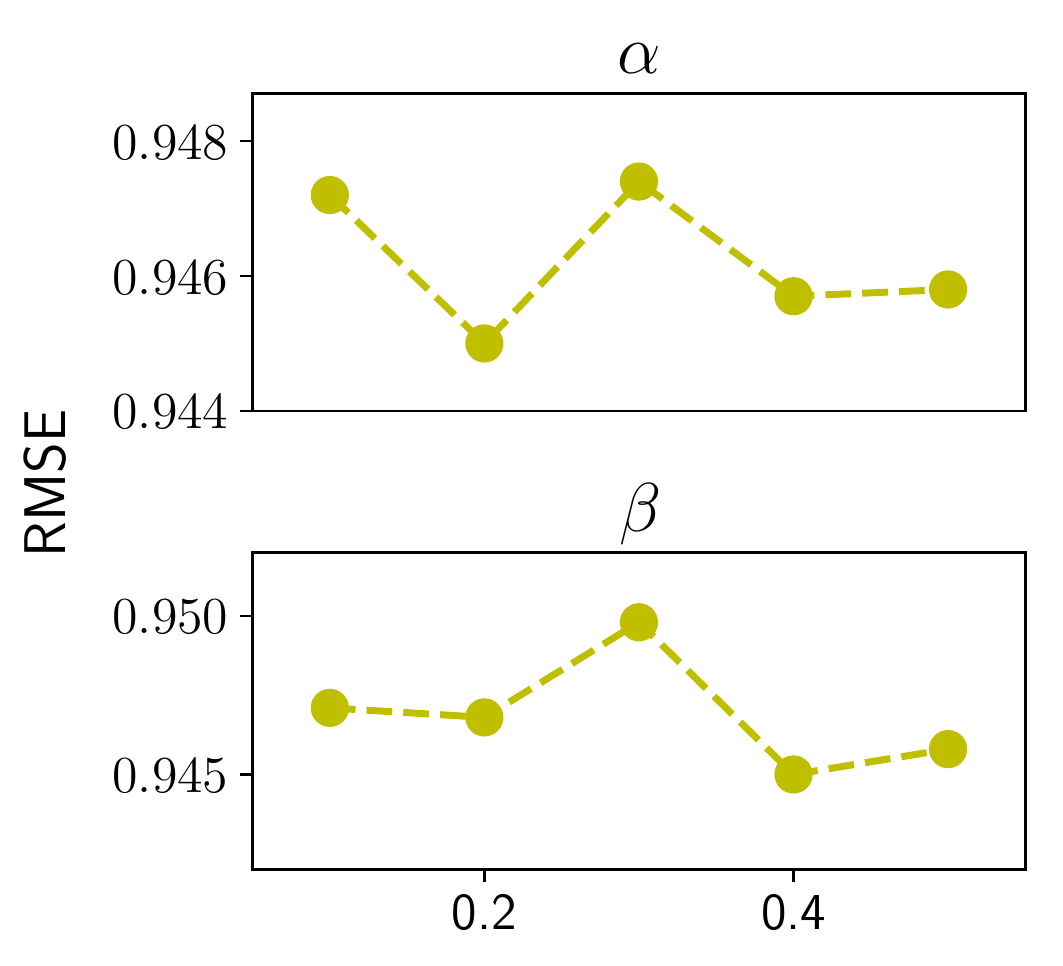}}
		\subfigure[Performance with $\gamma$ and $\theta$]{\label{fig:intro_2}\includegraphics[scale=0.37]{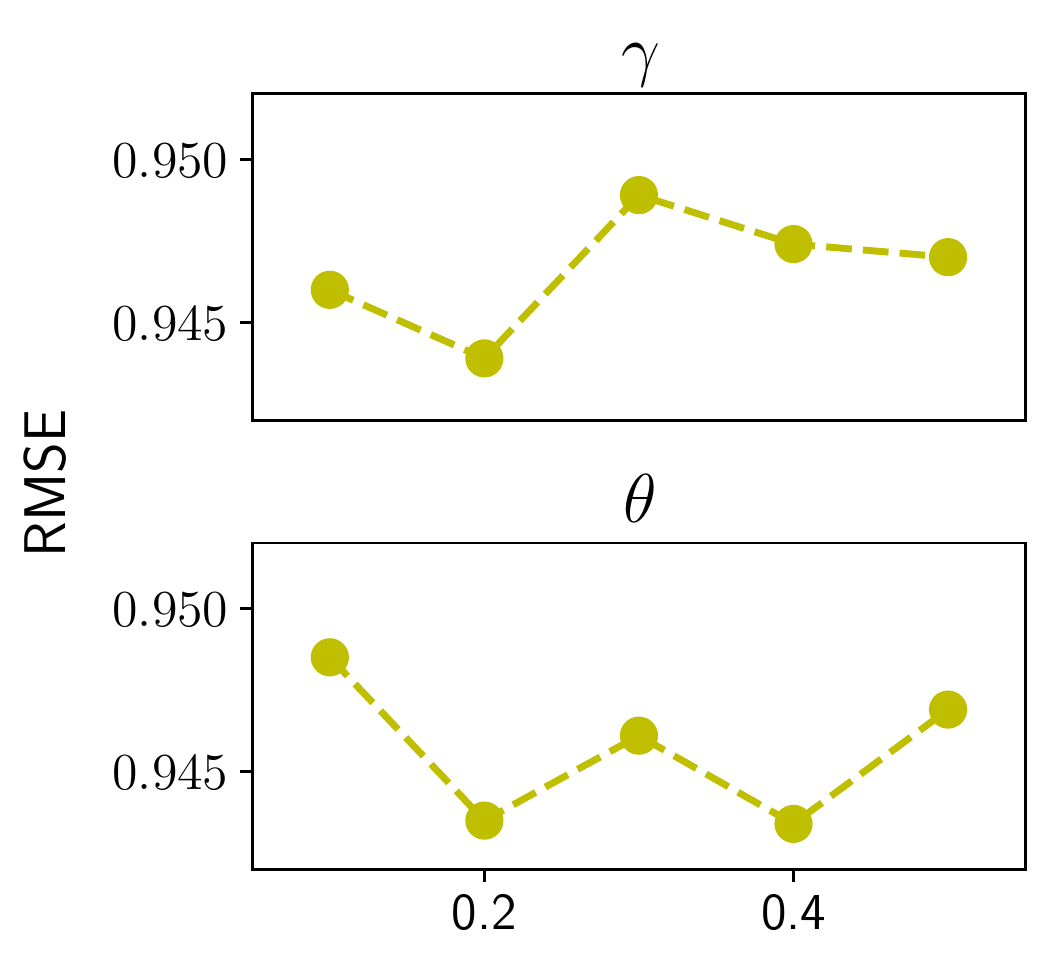}}
	\end{center}
	\caption{Parameter analysis with $\alpha$, $\beta$, $\gamma$ and $\theta$}
	\label{fig:para}
\end{figure}

In this subsection, we further analyze the sensitivity of the four key parameters $\gamma$, $\theta$, $\alpha$ and $\beta$ that control the contributions of flat side information of users, flat side information of items, hierarchical side information of users, and hierarchical information of items, respectively. In detail, for each of the four parameters, we conduct experiment with the proposed framework by varying the value of it while fixing the others. The performance is shown in Figure~\ref{fig:para}. Similarly, only performance on MovieLens~(100K) is reported due to the space limitation. From both figures, we clearly see that the performance tends to first increase and then decrease, which further supports the importance of side information in recommendations. 

\section{Related Work}
In this section, we give a brief overview of the related recommender systems. A large body of research has been devoted to developing algorithms to improve the performance of recommender systems, which play a crucial role in the increasingly digitalized society. Among them, collaborative filtering based approaches have achieved great success. Roughly, collaborative filtering can be categorized into two type: (1) memory-based approaches~\cite{sarwar2001item,wang2006unifying,melville2002content,popescul2001probabilistic}, which aim at exploring neighborhood information of users or items for recommendation; and (2) model-based methods~\cite{koren2009matrix,ma2008sorec,gu2010collaborative}, which try to model the underlying mechanism that governs user rating process. Generally, model-based methods show superior performance than the content-based ones. In particular, Matrix Factorization (MF) based collaborative filtering have gained great popularity due to their high performance and efficiency~\cite{lee1999learning,koren2009matrix,mnih2008probabilistic,salakhutdinov2008bayesian,srebro2005maximum}. Despite of its success, collaborative filtering approaches are known to suffer from data sparsity issues, as the number of items or users is typically very large but the number of ratings is relatively small. One popular way to address this issue is to incorporate the increasingly available side information in the model~\cite{fang2011matrix,vasile2016meta,tang2016recommendation,adomavicius2011context,wang2018exploring,lu2012exploiting}. The majority of studies exploit either only flat side information~\cite{fang2011matrix,adomavicius2011context}, or only hierarchical side information~\cite{wang2018exploring,lu2012exploiting} due to the challenges brought by the inherent difference between these two types of information. However, our work addresses these challenges and is able to incorporate the two types of information simultaneously.
\section{Conclusion}
In this paper, we investigate the problem of exploiting heterogeneous side information for recommendations. Specifically, we propose a novel recommendation framework HIRE that is able to capture both flat and hierarchical side information with mathematical coherence. Extensive experiments on three benchmark datasets verify the effectiveness of the framework and demonstrate the impact of both flat and hierarchical side information on recommendation performance. 

\section*{ACKNOWLEDGMENTS}
Jiliang Tang is supported by the National Science Foundation (NSF) under grant numbers IIS-1714741, IIS-1715940 and CNS-1815636, and a grant from Criteo Faculty Research Award.

\bibliographystyle{ACM-Reference-Format}
\bibliography{zhiwei}


\begin{thebibliography}{34}


\ifx \showCODEN    \undefined \def \showCODEN     #1{\unskip}     \fi
\ifx \showDOI      \undefined \def \showDOI       #1{#1}\fi
\ifx \showISBNx    \undefined \def \showISBNx     #1{\unskip}     \fi
\ifx \showISBNxiii \undefined \def \showISBNxiii  #1{\unskip}     \fi
\ifx \showISSN     \undefined \def \showISSN      #1{\unskip}     \fi
\ifx \showLCCN     \undefined \def \showLCCN      #1{\unskip}     \fi
\ifx \shownote     \undefined \def \shownote      #1{#1}          \fi
\ifx \showarticletitle \undefined \def \showarticletitle #1{#1}   \fi
\ifx \showURL      \undefined \def \showURL       {\relax}        \fi
\providecommand\bibfield[2]{#2}
\providecommand\bibinfo[2]{#2}
\providecommand\natexlab[1]{#1}
\providecommand\showeprint[2][]{arXiv:#2}

\bibitem[\protect\citeauthoryear{Adomavicius and Tuzhilin}{Adomavicius and
  Tuzhilin}{2005}]%
        {adomavicius2005toward}
\bibfield{author}{\bibinfo{person}{Gediminas Adomavicius} {and}
  \bibinfo{person}{Alexander Tuzhilin}.} \bibinfo{year}{2005}\natexlab{}.
\newblock \showarticletitle{Toward the next generation of recommender systems:
  A survey of the state-of-the-art and possible extensions}.
\newblock \bibinfo{journal}{\emph{TKDE}} \bibinfo{number}{6}
  (\bibinfo{year}{2005}), \bibinfo{pages}{734--749}.
\newblock


\bibitem[\protect\citeauthoryear{Adomavicius and Tuzhilin}{Adomavicius and
  Tuzhilin}{2011}]%
        {adomavicius2011context}
\bibfield{author}{\bibinfo{person}{Gediminas Adomavicius} {and}
  \bibinfo{person}{Alexander Tuzhilin}.} \bibinfo{year}{2011}\natexlab{}.
\newblock \showarticletitle{Context-aware recommender systems}.
\newblock In \bibinfo{booktitle}{\emph{Recommender systems handbook}}.
  \bibinfo{publisher}{Springer}, \bibinfo{pages}{217--253}.
\newblock


\bibitem[\protect\citeauthoryear{Bengio, Courville, and Vincent}{Bengio
  et~al\mbox{.}}{2013}]%
        {bengio2013representation}
\bibfield{author}{\bibinfo{person}{Yoshua Bengio}, \bibinfo{person}{Aaron
  Courville}, {and} \bibinfo{person}{Pascal Vincent}.}
  \bibinfo{year}{2013}\natexlab{}.
\newblock \showarticletitle{Representation learning: A review and new
  perspectives}.
\newblock \bibinfo{journal}{\emph{IEEE transactions on pattern analysis and
  machine intelligence}} \bibinfo{volume}{35}, \bibinfo{number}{8}
  (\bibinfo{year}{2013}), \bibinfo{pages}{1798--1828}.
\newblock


\bibitem[\protect\citeauthoryear{Chen, Xu, Weinberger, and Sha}{Chen
  et~al\mbox{.}}{2012}]%
        {chen2012marginalized}
\bibfield{author}{\bibinfo{person}{Minmin Chen}, \bibinfo{person}{Zhixiang Xu},
  \bibinfo{person}{Kilian Weinberger}, {and} \bibinfo{person}{Fei Sha}.}
  \bibinfo{year}{2012}\natexlab{}.
\newblock \showarticletitle{Marginalized denoising autoencoders for domain
  adaptation}.
\newblock \bibinfo{journal}{\emph{arXiv preprint arXiv:1206.4683}}
  (\bibinfo{year}{2012}).
\newblock


\bibitem[\protect\citeauthoryear{Fang and Si}{Fang and Si}{2011}]%
        {fang2011matrix}
\bibfield{author}{\bibinfo{person}{Yi Fang} {and} \bibinfo{person}{Luo Si}.}
  \bibinfo{year}{2011}\natexlab{}.
\newblock \showarticletitle{Matrix co-factorization for recommendation with
  rich side information and implicit feedback}. In
  \bibinfo{booktitle}{\emph{Proceedings of the 2nd International Workshop on
  Information Heterogeneity and Fusion in Recommender Systems}}. ACM,
  \bibinfo{pages}{65--69}.
\newblock


\bibitem[\protect\citeauthoryear{Golub and Reinsch}{Golub and Reinsch}{1970}]%
        {golub1970singular}
\bibfield{author}{\bibinfo{person}{Gene~H Golub} {and}
  \bibinfo{person}{Christian Reinsch}.} \bibinfo{year}{1970}\natexlab{}.
\newblock \showarticletitle{Singular value decomposition and least squares
  solutions}.
\newblock \bibinfo{journal}{\emph{Numerische mathematik}} \bibinfo{volume}{14},
  \bibinfo{number}{5} (\bibinfo{year}{1970}), \bibinfo{pages}{403--420}.
\newblock


\bibitem[\protect\citeauthoryear{Gu, Zhou, and Ding}{Gu et~al\mbox{.}}{2010}]%
        {gu2010collaborative}
\bibfield{author}{\bibinfo{person}{Quanquan Gu}, \bibinfo{person}{Jie Zhou},
  {and} \bibinfo{person}{Chris Ding}.} \bibinfo{year}{2010}\natexlab{}.
\newblock \showarticletitle{Collaborative filtering: Weighted nonnegative
  matrix factorization incorporating user and item graphs}. In
  \bibinfo{booktitle}{\emph{SDM}}. SIAM, \bibinfo{pages}{199--210}.
\newblock


\bibitem[\protect\citeauthoryear{Harper and Konstan}{Harper and
  Konstan}{2016}]%
        {harper2016movielens}
\bibfield{author}{\bibinfo{person}{F~Maxwell Harper} {and}
  \bibinfo{person}{Joseph~A Konstan}.} \bibinfo{year}{2016}\natexlab{}.
\newblock \showarticletitle{The movielens datasets: History and context}.
\newblock \bibinfo{journal}{\emph{Acm transactions on interactive intelligent
  systems (tiis)}} \bibinfo{volume}{5}, \bibinfo{number}{4}
  (\bibinfo{year}{2016}), \bibinfo{pages}{19}.
\newblock


\bibitem[\protect\citeauthoryear{He, Liao, Zhang, Nie, Hu, and Chua}{He
  et~al\mbox{.}}{2017}]%
        {he2017neural}
\bibfield{author}{\bibinfo{person}{Xiangnan He}, \bibinfo{person}{Lizi Liao},
  \bibinfo{person}{Hanwang Zhang}, \bibinfo{person}{Liqiang Nie},
  \bibinfo{person}{Xia Hu}, {and} \bibinfo{person}{Tat-Seng Chua}.}
  \bibinfo{year}{2017}\natexlab{}.
\newblock \showarticletitle{Neural collaborative filtering}. In
  \bibinfo{booktitle}{\emph{WWW}}. \bibinfo{pages}{173--182}.
\newblock


\bibitem[\protect\citeauthoryear{Koren}{Koren}{2008}]%
        {koren2008factorization}
\bibfield{author}{\bibinfo{person}{Yehuda Koren}.}
  \bibinfo{year}{2008}\natexlab{}.
\newblock \showarticletitle{Factorization meets the neighborhood: a
  multifaceted collaborative filtering model}. In
  \bibinfo{booktitle}{\emph{KDD}}. ACM, \bibinfo{pages}{426--434}.
\newblock


\bibitem[\protect\citeauthoryear{Koren, Bell, and Volinsky}{Koren
  et~al\mbox{.}}{2009}]%
        {koren2009matrix}
\bibfield{author}{\bibinfo{person}{Yehuda Koren}, \bibinfo{person}{Robert
  Bell}, {and} \bibinfo{person}{Chris Volinsky}.}
  \bibinfo{year}{2009}\natexlab{}.
\newblock \showarticletitle{Matrix factorization techniques for recommender
  systems}.
\newblock \bibinfo{journal}{\emph{Computer}} \bibinfo{number}{8}
  (\bibinfo{year}{2009}), \bibinfo{pages}{30--37}.
\newblock


\bibitem[\protect\citeauthoryear{Lee and Seung}{Lee and Seung}{1999}]%
        {lee1999learning}
\bibfield{author}{\bibinfo{person}{Daniel~D Lee} {and}
  \bibinfo{person}{H~Sebastian Seung}.} \bibinfo{year}{1999}\natexlab{}.
\newblock \showarticletitle{Learning the parts of objects by non-negative
  matrix factorization}.
\newblock \bibinfo{journal}{\emph{Nature}} \bibinfo{volume}{401},
  \bibinfo{number}{6755} (\bibinfo{year}{1999}), \bibinfo{pages}{788}.
\newblock


\bibitem[\protect\citeauthoryear{Li, Luong, and Jurafsky}{Li
  et~al\mbox{.}}{2015b}]%
        {li2015hierarchical}
\bibfield{author}{\bibinfo{person}{Jiwei Li}, \bibinfo{person}{Minh-Thang
  Luong}, {and} \bibinfo{person}{Dan Jurafsky}.}
  \bibinfo{year}{2015}\natexlab{b}.
\newblock \showarticletitle{A hierarchical neural autoencoder for paragraphs
  and documents}.
\newblock \bibinfo{journal}{\emph{arXiv preprint arXiv:1506.01057}}
  (\bibinfo{year}{2015}).
\newblock


\bibitem[\protect\citeauthoryear{Li, Kawale, and Fu}{Li et~al\mbox{.}}{2015a}]%
        {li2015deep}
\bibfield{author}{\bibinfo{person}{Sheng Li}, \bibinfo{person}{Jaya Kawale},
  {and} \bibinfo{person}{Yun Fu}.} \bibinfo{year}{2015}\natexlab{a}.
\newblock \showarticletitle{Deep collaborative filtering via marginalized
  denoising auto-encoder}. In \bibinfo{booktitle}{\emph{CIKM}}. ACM,
  \bibinfo{pages}{811--820}.
\newblock


\bibitem[\protect\citeauthoryear{Lu, Zhang, Li, Zhang, and Wang}{Lu
  et~al\mbox{.}}{2012}]%
        {lu2012exploiting}
\bibfield{author}{\bibinfo{person}{Kai Lu}, \bibinfo{person}{Guanyuan Zhang},
  \bibinfo{person}{Rui Li}, \bibinfo{person}{Shuai Zhang}, {and}
  \bibinfo{person}{Bin Wang}.} \bibinfo{year}{2012}\natexlab{}.
\newblock \showarticletitle{Exploiting and exploring hierarchical structure in
  music recommendation}. In \bibinfo{booktitle}{\emph{Asia Information
  Retrieval Symposium}}. Springer, \bibinfo{pages}{211--225}.
\newblock


\bibitem[\protect\citeauthoryear{Lu, Tsao, Matsuda, and Hori}{Lu
  et~al\mbox{.}}{2013}]%
        {lu2013speech}
\bibfield{author}{\bibinfo{person}{Xugang Lu}, \bibinfo{person}{Yu Tsao},
  \bibinfo{person}{Shigeki Matsuda}, {and} \bibinfo{person}{Chiori Hori}.}
  \bibinfo{year}{2013}\natexlab{}.
\newblock \showarticletitle{Speech enhancement based on deep denoising
  autoencoder.}. In \bibinfo{booktitle}{\emph{Interspeech}}.
  \bibinfo{pages}{436--440}.
\newblock


\bibitem[\protect\citeauthoryear{Ma, Yang, Lyu, and King}{Ma
  et~al\mbox{.}}{2008}]%
        {ma2008sorec}
\bibfield{author}{\bibinfo{person}{Hao Ma}, \bibinfo{person}{Haixuan Yang},
  \bibinfo{person}{Michael~R Lyu}, {and} \bibinfo{person}{Irwin King}.}
  \bibinfo{year}{2008}\natexlab{}.
\newblock \showarticletitle{Sorec: social recommendation using probabilistic
  matrix factorization}. In \bibinfo{booktitle}{\emph{CIKM}}. ACM,
  \bibinfo{pages}{931--940}.
\newblock


\bibitem[\protect\citeauthoryear{Melville, Mooney, and Nagarajan}{Melville
  et~al\mbox{.}}{2002}]%
        {melville2002content}
\bibfield{author}{\bibinfo{person}{Prem Melville}, \bibinfo{person}{Raymond~J
  Mooney}, {and} \bibinfo{person}{Ramadass Nagarajan}.}
  \bibinfo{year}{2002}\natexlab{}.
\newblock \showarticletitle{Content-boosted collaborative filtering for
  improved recommendations}.
\newblock \bibinfo{journal}{\emph{Aaai/iaai}}  \bibinfo{volume}{23}
  (\bibinfo{year}{2002}), \bibinfo{pages}{187--192}.
\newblock


\bibitem[\protect\citeauthoryear{Menon, Chitrapura, Garg, Agarwal, and
  Kota}{Menon et~al\mbox{.}}{2011}]%
        {menon2011response}
\bibfield{author}{\bibinfo{person}{Aditya~Krishna Menon},
  \bibinfo{person}{Krishna-Prasad Chitrapura}, \bibinfo{person}{Sachin Garg},
  \bibinfo{person}{Deepak Agarwal}, {and} \bibinfo{person}{Nagaraj Kota}.}
  \bibinfo{year}{2011}\natexlab{}.
\newblock \showarticletitle{Response prediction using collaborative filtering
  with hierarchies and side-information}. In
  \bibinfo{booktitle}{\emph{Proceedings of the 17th ACM SIGKDD international
  conference on Knowledge discovery and data mining}}. ACM,
  \bibinfo{pages}{141--149}.
\newblock


\bibitem[\protect\citeauthoryear{Mnih and Salakhutdinov}{Mnih and
  Salakhutdinov}{2008}]%
        {mnih2008probabilistic}
\bibfield{author}{\bibinfo{person}{Andriy Mnih} {and} \bibinfo{person}{Ruslan~R
  Salakhutdinov}.} \bibinfo{year}{2008}\natexlab{}.
\newblock \showarticletitle{Probabilistic matrix factorization}. In
  \bibinfo{booktitle}{\emph{NIPS}}. \bibinfo{pages}{1257--1264}.
\newblock


\bibitem[\protect\citeauthoryear{Ning and Karypis}{Ning and Karypis}{2012}]%
        {ning2012sparse}
\bibfield{author}{\bibinfo{person}{Xia Ning} {and} \bibinfo{person}{George
  Karypis}.} \bibinfo{year}{2012}\natexlab{}.
\newblock \showarticletitle{Sparse linear methods with side information for
  top-n recommendations}. In \bibinfo{booktitle}{\emph{RecSys}}. ACM,
  \bibinfo{pages}{155--162}.
\newblock


\bibitem[\protect\citeauthoryear{Popescul, Pennock, and Lawrence}{Popescul
  et~al\mbox{.}}{2001}]%
        {popescul2001probabilistic}
\bibfield{author}{\bibinfo{person}{Alexandrin Popescul},
  \bibinfo{person}{David~M Pennock}, {and} \bibinfo{person}{Steve Lawrence}.}
  \bibinfo{year}{2001}\natexlab{}.
\newblock \showarticletitle{Probabilistic models for unified collaborative and
  content-based recommendation in sparse-data environments}. In
  \bibinfo{booktitle}{\emph{UAI}}. Morgan Kaufmann Publishers Inc.,
  \bibinfo{pages}{437--444}.
\newblock


\bibitem[\protect\citeauthoryear{Ricci, Rokach, and Shapira}{Ricci
  et~al\mbox{.}}{2015}]%
        {ricci2015recommender}
\bibfield{author}{\bibinfo{person}{Francesco Ricci}, \bibinfo{person}{Lior
  Rokach}, {and} \bibinfo{person}{Bracha Shapira}.}
  \bibinfo{year}{2015}\natexlab{}.
\newblock \showarticletitle{Recommender systems: introduction and challenges}.
\newblock In \bibinfo{booktitle}{\emph{Recommender systems handbook}}.
  \bibinfo{publisher}{Springer}, \bibinfo{pages}{1--34}.
\newblock


\bibitem[\protect\citeauthoryear{Salakhutdinov and Mnih}{Salakhutdinov and
  Mnih}{2008}]%
        {salakhutdinov2008bayesian}
\bibfield{author}{\bibinfo{person}{Ruslan Salakhutdinov} {and}
  \bibinfo{person}{Andriy Mnih}.} \bibinfo{year}{2008}\natexlab{}.
\newblock \showarticletitle{Bayesian probabilistic matrix factorization using
  Markov chain Monte Carlo}. In \bibinfo{booktitle}{\emph{ICML}}. ACM,
  \bibinfo{pages}{880--887}.
\newblock


\bibitem[\protect\citeauthoryear{Sarwar, Karypis, Konstan, and Riedl}{Sarwar
  et~al\mbox{.}}{2001}]%
        {sarwar2001item}
\bibfield{author}{\bibinfo{person}{Badrul Sarwar}, \bibinfo{person}{George
  Karypis}, \bibinfo{person}{Joseph Konstan}, {and} \bibinfo{person}{John
  Riedl}.} \bibinfo{year}{2001}\natexlab{}.
\newblock \showarticletitle{Item-based collaborative filtering recommendation
  algorithms}. In \bibinfo{booktitle}{\emph{WWW}}. ACM,
  \bibinfo{pages}{285--295}.
\newblock


\bibitem[\protect\citeauthoryear{Srebro, Rennie, and Jaakkola}{Srebro
  et~al\mbox{.}}{2005}]%
        {srebro2005maximum}
\bibfield{author}{\bibinfo{person}{Nathan Srebro}, \bibinfo{person}{Jason
  Rennie}, {and} \bibinfo{person}{Tommi~S Jaakkola}.}
  \bibinfo{year}{2005}\natexlab{}.
\newblock \showarticletitle{Maximum-margin matrix factorization}. In
  \bibinfo{booktitle}{\emph{NIPS}}. \bibinfo{pages}{1329--1336}.
\newblock


\bibitem[\protect\citeauthoryear{Su and Khoshgoftaar}{Su and
  Khoshgoftaar}{2009}]%
        {su2009survey}
\bibfield{author}{\bibinfo{person}{Xiaoyuan Su} {and} \bibinfo{person}{Taghi~M
  Khoshgoftaar}.} \bibinfo{year}{2009}\natexlab{}.
\newblock \showarticletitle{A survey of collaborative filtering techniques}.
\newblock \bibinfo{journal}{\emph{Advances in artificial intelligence}}
  \bibinfo{volume}{2009} (\bibinfo{year}{2009}).
\newblock


\bibitem[\protect\citeauthoryear{Tang, Wang, Hu, Yin, Bi, Chang, and Liu}{Tang
  et~al\mbox{.}}{2016}]%
        {tang2016recommendation}
\bibfield{author}{\bibinfo{person}{Jiliang Tang}, \bibinfo{person}{Suhang
  Wang}, \bibinfo{person}{Xia Hu}, \bibinfo{person}{Dawei Yin},
  \bibinfo{person}{Yingzhou Bi}, \bibinfo{person}{Yi Chang}, {and}
  \bibinfo{person}{Huan Liu}.} \bibinfo{year}{2016}\natexlab{}.
\newblock \showarticletitle{Recommendation with Social Dimensions.}. In
  \bibinfo{booktitle}{\emph{AAAI}}. \bibinfo{pages}{251--257}.
\newblock


\bibitem[\protect\citeauthoryear{Vasile, Smirnova, and Conneau}{Vasile
  et~al\mbox{.}}{2016}]%
        {vasile2016meta}
\bibfield{author}{\bibinfo{person}{Flavian Vasile}, \bibinfo{person}{Elena
  Smirnova}, {and} \bibinfo{person}{Alexis Conneau}.}
  \bibinfo{year}{2016}\natexlab{}.
\newblock \showarticletitle{Meta-prod2vec: Product embeddings using
  side-information for recommendation}. In \bibinfo{booktitle}{\emph{RecSys}}.
  ACM, \bibinfo{pages}{225--232}.
\newblock


\bibitem[\protect\citeauthoryear{Wang, De~Vries, and Reinders}{Wang
  et~al\mbox{.}}{2006}]%
        {wang2006unifying}
\bibfield{author}{\bibinfo{person}{Jun Wang}, \bibinfo{person}{Arjen~P
  De~Vries}, {and} \bibinfo{person}{Marcel~JT Reinders}.}
  \bibinfo{year}{2006}\natexlab{}.
\newblock \showarticletitle{Unifying user-based and item-based collaborative
  filtering approaches by similarity fusion}. In
  \bibinfo{booktitle}{\emph{SIGIR}}. ACM, \bibinfo{pages}{501--508}.
\newblock


\bibitem[\protect\citeauthoryear{Wang, Tang, Wang, and Liu}{Wang
  et~al\mbox{.}}{2018}]%
        {wang2018exploring}
\bibfield{author}{\bibinfo{person}{Suhang Wang}, \bibinfo{person}{Jiliang
  Tang}, \bibinfo{person}{Yilin Wang}, {and} \bibinfo{person}{Huan Liu}.}
  \bibinfo{year}{2018}\natexlab{}.
\newblock \showarticletitle{Exploring Hierarchical Structures for Recommender
  Systems}.
\newblock \bibinfo{journal}{\emph{TKDE}} \bibinfo{volume}{30},
  \bibinfo{number}{6} (\bibinfo{year}{2018}), \bibinfo{pages}{1022--1035}.
\newblock


\bibitem[\protect\citeauthoryear{Yang, Sun, Bozzon, and Zhang}{Yang
  et~al\mbox{.}}{2016}]%
        {yang2016learning}
\bibfield{author}{\bibinfo{person}{Jie Yang}, \bibinfo{person}{Zhu Sun},
  \bibinfo{person}{Alessandro Bozzon}, {and} \bibinfo{person}{Jie Zhang}.}
  \bibinfo{year}{2016}\natexlab{}.
\newblock \showarticletitle{Learning hierarchical feature influence for
  recommendation by recursive regularization}. In
  \bibinfo{booktitle}{\emph{RecSys}}. ACM, \bibinfo{pages}{51--58}.
\newblock


\bibitem[\protect\citeauthoryear{Zhang, Shan, Kan, and Chen}{Zhang
  et~al\mbox{.}}{2014}]%
        {zhang2014coarse}
\bibfield{author}{\bibinfo{person}{Jie Zhang}, \bibinfo{person}{Shiguang Shan},
  \bibinfo{person}{Meina Kan}, {and} \bibinfo{person}{Xilin Chen}.}
  \bibinfo{year}{2014}\natexlab{}.
\newblock \showarticletitle{Coarse-to-fine auto-encoder networks (cfan) for
  real-time face alignment}. In \bibinfo{booktitle}{\emph{ECCV}}. Springer,
  \bibinfo{pages}{1--16}.
\newblock


\bibitem[\protect\citeauthoryear{Ziegler, McNee, Konstan, and Lausen}{Ziegler
  et~al\mbox{.}}{2005}]%
        {ziegler2005improving}
\bibfield{author}{\bibinfo{person}{Cai-Nicolas Ziegler},
  \bibinfo{person}{Sean~M McNee}, \bibinfo{person}{Joseph~A Konstan}, {and}
  \bibinfo{person}{Georg Lausen}.} \bibinfo{year}{2005}\natexlab{}.
\newblock \showarticletitle{Improving recommendation lists through topic
  diversification}. In \bibinfo{booktitle}{\emph{WWW}}. ACM,
  \bibinfo{pages}{22--32}.
\newblock


\end{thebibliography}

\end{document}